\documentclass[twocolumn, nofootinbib,nobibnotes, showpacs]{revtex4}
\usepackage{amsmath,amssymb,amsfonts,times,graphicx,color}
\usepackage{esint}

\usepackage{hyperref}
\usepackage{eucal}
\definecolor{darkred}{rgb}{0.8,0.1,0.1}
\hypersetup{colorlinks=true, linkcolor=darkred, citecolor=blue}

\begin{document}

\title{Excited state entanglement in one dimensional quantum critical systems:\\
extensivity and the role of microscopic details}

\author{T. P\'almai}
\email{tpalmai@sissa.it}

\affiliation{\textit{\small Scuola Internazionale Superiore di Studi Avanzati
Trieste}{\small{} }\\
{\small{} }\textit{\small Istituto Nazionale di Fisica Nucleare, Sezione
Trieste}{\small{} }\\
{\small{} }\textit{\small Via Bonomea 265, 34136 Trieste, Italy}{\small{} }}
\begin{abstract}
We study entanglement via the subsystem purity relative to bipartitions
of arbitrary excited states in (1+1)-dimensional conformal field theory, equivalent
to the scaling limit of one dimensional quantum critical systems. We compute the exact subpurity 
as a function of the relative subsystem size for numerous excited states 
in the Ising and three-state Potts models. We find that it
decays exponentially when the system and the subsystem sizes are comparable
until a saturation limit is reached near half-partitioning, signaling that
excited states are maximally entangled. The exponential behavior translates
into extensivity for the second R\'enyi entropy. Since the coefficient of this linear law depends 
only on the excitation energy, this result shows an interesting, new relationship 
between energy and quantum information and elucidates the role of microscopic details.
\end{abstract}


\maketitle

Entanglement is the essence of quantum theory. Beyond quantum informational
aspects, the amount of entanglement coded into a quantum many-body
system is an increasingly important quantity, that provides a universal
way to characterize quantum fluctuation. The ground state entanglement
in particular can classify quantum phases, e.g it can 
inform us about interesting, topological phases \cite{topol} and whether a system is 
close to criticality \cite{AFOV}.

On the other hand, entanglement of excited states is also a very interesting quantity. 
In fact, recently there has been a quickly growing interest in this subject \cite{excited}.
While ground states characterize system specific quantum fluctuations and they usually
follow the area law \cite{arealaw}, the entanglement entropy of highly excited states 
are expected to be more generic, independent of the specific system and
go to an extensive, thermodynamic entropy instead.

This picture gets modified for critical systems. Here the area law for the ground state 
is enhanced into a logarithmic law \cite{HLW,CC} with a universal coefficient.
For low-lying excited states the logarithmic behavior was shown to remain unchanged when the subsystem size 
is infinitesimal compared to the whole. On the other hand, when the subsystem is 
comparable to the whole interesting features emerge, in particular hints of a thermodynamic behavior for very
small (but not infinitesimal) subsystem sizes, connecting energy with
the amount of entanglement in a universal way \cite{BNTU,Sierra}.
Despite these advances, there is still very little known about the entanglement
of excited states in critical systems, which however are especially interesting
because of universality. Interesting questions include: What happens for highly excited states?
How does entanglement behave for comparable subsystems? What is the relationship between entanglement 
and excitation energy beyond the small subsystem limit?

In this paper we endeavor to answer these questions by studying 
excited states in (1+1)-dimensional (unitary) conformal field theories. We set up a 
systematic framework to compute the R\'enyi entropies for {\it arbitrary} excited states relative to
connected bipartitions, by generalizing the approach of \cite{Sierra},
where the case of primary states was considered. The main motivation 
is that primary states represent only the lowest-lying excitations and in many important cases
(e.g. minimal models) there are only a few of them in the Hilbert space.
In order to examine more excited states, and in particular, to access highly 
excited ones, it is necessary to generalize the previous approach to 
descendant states.

We study in particular the scaling of the second R\'enyi entropy $S_2(d)$ 
(defined precisely below) with the relative subsystem size $d=L_A/L$ in the critical Ising and three state 
Potts models for several individual excited states. We chose the second R\'enyi entropy in addition to being 
computationally the simplest case, because it has the interpretation of being the logarithm
of the purity of the subsystem (defined as the participation ratio in the Schmidt basis, also see later).
We note that the choice of the particular models is arbitrary and one could choose any other minimal model, or in general a (1+1)-d CFT, where the four point functions of primary fields are known, and obtain the R\'enyi entropies by the present approach. 

We find that for almost all the excited states there are three distinct regimes
of the second R\'enyi entropy as a function of the relative subsystem 
size. The first regime has already 
been understood in earlier works \cite{Sierra,BNTU} and it corresponds to the small subsystem size limit 
described by a logarithmic scaling with the first {\it correction} being linear in the excitation energy
\begin{equation}\label{Ssmalld}
S_{2}^\Psi(d\to0)=\frac{c}{4}\log\sin\pi d+\frac{(\Delta^\Psi+\bar{\Delta}^\Psi)}{2}(\pi d)^{2}+\ldots
\end{equation}
with $c$ the central charge and $\Delta_\Psi$, $\bar{\Delta}_\Psi$ are the chiral and
antichiral scaling weights of the operator $\Psi$. (Here and in the rest of this paper
we do not write out additive constants common to all states, e.g. the UV regularization factor.) 
Although this form was proved rigorously only for primary states it is expected to hold for 
other excited states as well \cite{BNTU}, and we could confirm this for the states that we considered.
Beyond this regime no general feature was known until now.
%Indeed, one can argue that the leading logarithmic 
%behavior is general for any excited state, however the thermodynamic correction is not apparent from the 
%present formulation.

The second regime is the most interesting and its identification is the main result of this Letter. It is characterized by an extensive R\'enyi entropy, growing linearly with the subsystem size
\begin{equation}
S_{2}^{\Psi}(d)\approx s(E^{\Psi})d,\qquad d_{\text{log.}}<d<d_{\text{sat.}}\label{eq:Sexc}
\end{equation}
The scaling law only depends on the excitation energy and it is {\it sublinear} in it.
In other words (in the accessible energy range) we see a quasi-thermodynamic entropy.
Extensivity translates for the subsystem purity as exponential decay, signaling 
that the excited states are maximally entangled. We refer to \cite{holoext} 
where extensive terms were observed for special states in the entanglement entropy in holographic systems.

There is a third regime, which turns out to be unique to every state and it corresponds
to the saturation of the entropy to facilitate the symmetry $S(d)=S(1-d)$. This
part of the R\'enyi entropy can be used to differentiate between degenerate states
and allows for instance the identification of excited states of the microscopic theory
in terms of CFT states (suggested first in \cite{Sierra}).

In our explicit calculations we determined that these regimes can be found for all 
(primary and descendant) states in the accessible energy range, except for the 
lowest-lying excitation in the Ising CFT corresponding to the twisted vacuum in the 
spin chain model. Therefore, it is reasonable to believe that these features constitute 
a general property of the second Renyi entropy (subsystem purity) for most of the excited 
states. In fact, the only states for which Eqs. (\ref{Ssmalld}-\ref{eq:Sexc}) are expected 
to break down are those corresponding to ground states of some local Hamiltonians 
(see \cite{exceptions} and also \cite{excspinchain} for an analysis in the XY spin chain).

In addition to the present results we expect that generalizing the
computation of R\'enyi entropies to arbitrary states in CFT will
enable a number of exciting future applications, e.g. the study 
of ground and excited state entanglement in one dimensional 
massive field theories (integrable or not) through the truncated conformal
space approach \cite{TCSA} or the study of local quenches \cite{localq}
in both gapless and gapful systems.

In the rest of this paper we first introduce the measure of entanglement
considered here and then outline the computation technique to find 
the exact R\'enyi entropies and in particular,
the second one equivalent to the subsystem purity, for arbitrary excited
states. We also present representative results for excited states
in the Ising and three state Potts universality classes.

\paragraph*{R\'enyi entropies in CFT}

To define a measure of entanglement for a system in a pure state it
is enough to look at a spatial bipartition ($A\cup B$). A family
of measures coming from the reduced density matrix on $A$ consists of the
R\'enyi entropies, defined as
\begin{equation}
S_{n}=\frac{1}{1-n}\log\text{Tr}_{A}\rho_{A}^{n},\qquad\rho_{A}=\text{Tr}_{B}\vert\Psi\rangle\langle\Psi\vert,\label{eq:Sn}
\end{equation}
Beside the von Neumann entropy $S_{1}=\lim_{n\to1}S_{n}$, also
especially important is the second R\'enyi entropy that measures the
purity of the subsystem. The concept of purity can be introduced 
considering the Schmidt decomposition,
\begin{equation}
\vert\Psi\rangle=\sum_m c_m\vert a_m\rangle \vert b_m\rangle
\end{equation}
in terms of these bases relative to the partitions the reduced
density matrix takes the form
\begin{equation}
\rho_{A}=\sum_{m}c_{m}^2\vert a_m\rangle\langle a_m\vert
\end{equation}
where we can see, that the coherences are zero and in this sense
the basis involved in the Schmidt decomposition is a maximally entangled
basis. The participation ratio $P=\sum_{m}c_{m}^{4}=\text{Tr}_{A}\rho_{A}^{2}$
on this basis then describes the purity of the subsystem. If the subsystem
can be described by a pure state purity would be one, while for mixed
states it would give the inverse of the effective number of maximally
entangled states needed to describe it. The more this number goes
to zero the less pure the mixed state is, therefore the more the partitions
are entangled. Since the dimension of the subsystem Hilbert space
scales exponentially with its size, exponential decay (i.e. extensivity
of the R\'enyi entropy) would mean maximally entangled states.

Turning now to the computation of this subpurity
in our field theoretical setting we expand the traces in (\ref{eq:Sn}):
the relevant expression can be written in terms of sums on two bases
relative to $A$ and $B$ (e.g. but not necessarily the Schmidt bases). Since we are in finite volume
the energy levels are quantized and we can write
\begin{eqnarray}
P=\text{Tr}_{A}\rho_{A}^{2} & = & \sum_{a}\langle a\vert\left(\sum_{b}\langle b\vert\vert\Psi\rangle\langle\Psi\vert\vert b\rangle\right)^{2}\vert a\rangle\\
 & = & \sum_{aa'bb'}\langle ab\vert\Psi\rangle\langle\Psi\vert a'b\rangle\langle a'b'\vert\Psi\rangle\langle\Psi\vert ab'\rangle
\end{eqnarray}
where states labeled by $a$, $a'$ live on the partition $A$, while
$b$, $b'$ on $B$. The main problem is that a priori it is not clear
how to obtain the sets of states living on restrictions in terms of
those living on the full domain. However, one can reinterpret the
sum by noticing, that it is equivalent to a four point function 
(first noticed in \cite{Sierra}) on
a non-trivial geometry, that consists of two sheets (with periodic
boundary conditions on each) sewed together along $A$ in a circular
manner (for a pictorial representation see e.g. \cite{CC} where this
surface is denoted by $\mathcal{R}_2$). 

It is useful to go from interconnected cylinders ($\mathcal{R}_2$) 
to interconnected planes ($\bar{\mathcal{R}}_2$) by the 
exponential mapping $\xi=e^{-\frac{2\pi i}{L}(x+it)}$, where the physical energy
eigenstates are generated by the insertion of the usual primary and descendant
fields. The purity is then just the unusually normalized four-point function
\begin{equation}
P=\mathcal{N}_\Psi F_\Psi=
\mathcal{N}_\Psi\langle\Psi(0_{1})\Psi(0_{1})^{\dagger}\Psi(0_{2})\Psi(0_{2})^{\dagger}\rangle_{\bar{\mathcal{R}}_{2}}
\end{equation}
where the operator $\Psi$ implements the state $\vert\Psi\rangle$
on the complex plane as
\begin{equation}
\Psi(0)\vert0\rangle=\vert\Psi\rangle
\end{equation}
and the normalization $\mathcal{N}_\Psi$ is such that $\text{Tr}\rho=1$.

To evaluate $P$ we map $\bar{\mathcal{R}}_2$ to a single complex plane by the transformation 
\begin{equation}\label{trafo}
\zeta=f_{d}(\xi)=\left(\frac{e^{i\pi d}\xi-e^{-i\pi d}}{1-\xi}\right)^{1/2}
\end{equation}
The above mapping consists of the composition
of a global conformal mapping (Moebius transformation) and taking the square root.
The first component maps to a surface topologically identical to the Riemann surface of the square root,
being two sheeted with the connecting branch cut on the positive real line. Taking the
square root then maps to the complex plane, and the two sheets are prescribed to be mapped to the two branches of
the complex square root, e.g. $\xi=0_{1,2}$ maps to $e^{\frac{i\pi}{2}(2m-1-d)}$,
$m=1,2$. It is important to perform UV and IR regularizations by
excluding the neighborhoods of the common points of $A$ and $B$
and going to finite volume, making i) the plane topologically compatible
with $\mathcal{R}_{2}$ and ii) the entropies finite. The UV regulator will
only appear in the normalization.

In case of the vacuum $\Psi\equiv1$ the non-trivial part is given 
by only the normalization
\begin{equation}
 \mathcal{N}_\Psi=\mathcal{N}_1=\frac{Z_{\bar{\mathcal{R}}_2}}{Z_{\bar{\mathcal{R}}_1}^2}=\left(\frac{L}{\pi\varepsilon}\sin\pi d\right)^{-\frac{c}{4}}
\end{equation}
with $c$ being the central charge, $L$ the IR and $\varepsilon$ the UV regulators. $\mathcal{N}_1$
gives rise to the ubiquitous logarithm law \cite{HLW,CC} for the ground state.
This formula can be seen simply by noticing that the transformed geometry
is equivalent to a finite cylinder where the partition function is well known.
It is also easy to see, that the leading logarithmic law for small subsystem sizes is 
universal and it is independent of which state is examined by 
considering that $\bar{\mathcal{R}}_2$ goes to two independent 
planes when the subsystem size goes to zero. Then, the four point
function gives $1$ because the states are normalized on the plane
and only $\mathcal{N}_\Psi$ remains.

When moving on to excited states one needs to evaluate a nontrivial $F_\Psi$. This
transforms under the same mapping $f_d$ into the equal-time 
($|\zeta|=1$ in radial quantization) four-point function 
\begin{equation}\label{Ftrafd}
F_\Psi=\langle\mathcal{T}\{\Psi(0_1)\}\mathcal{T}\{\Psi(0_1)^{\dagger}\}\mathcal{T}\{\Psi(0_2)\}\mathcal{T}\{\Psi(0_2)^{\dagger}\}\rangle_{\mathbb{C}}
\end{equation}
on the complex plane. $\mathcal{T}$ represents the transformation
of the fields under $f_{d}(\xi)$. For primary states the transformation
of the fields is simple,
\begin{equation}
 \mathcal{T}\{\Psi(0)\}=f_d'(0)^{2h+2\bar{h}}\Psi(f_d(0))
\end{equation}
with $h$ ($\bar{h}$) being the (anti)chiral primary weights. In this case
$F_\Psi$ is given by a four point function of the same operators
$\Psi$ that generate the physical state \cite{Sierra,SierraJStat},
\begin{equation}
F_\Psi\propto f_d'^{8h+8\bar{h}}\langle\Psi(\zeta_{1})\Psi(\zeta_{2})\Psi(\zeta_{3})\Psi(\zeta_{4})\rangle_{\mathbb{C}}
\end{equation}
with $\zeta_{1,2}=e^{\frac{i\pi\mp i\pi d}{2}}$, $\zeta_{3,4}=e^{\frac{3i\pi\mp i\pi d}{2}}$
and the only ambiguity is a phase factor that is set by $F_\Psi$ being positive real.

\paragraph*{Descendant states}

For descendants Eq. (\ref{Ftrafd}) must be evaluated much more carefully. 
In fact, in this case because of the generation of lower descendants when
performing the transformation for every descendant state there is a different
formula in terms of four point functions on the plane, which can in turn be
evaluated by standard methods.

To make progress we need to transform descendant operators under $f_{d}$. 
(Note that the images of the adjoints can be obtained from fields living in $\xi=0$ by $f_{-d}$.) 
The first term of the transformation law of a (chiral) descendant operator
(with the non-chiral case being a product of the chiral and antichiral
contributions) is easily obtained in general as
\begin{equation}
\mathcal{T}\{\phi(\zeta)\}=\left(\frac{\partial f}{\partial \zeta}\right)^{\Delta}\phi(f(\zeta))+\ldots
\end{equation}
for a field $\phi$ with scaling weight $\Delta$, however for a descendant
field all the lower descendants in the given tower are also generated
(represented above by ``$\ldots$'') and they can by no means be disregarded
in the four-point functions. To perform this transformation for general fields we propose
to use the construction of \cite{Gaberdiel} prescribing the form
\begin{equation}
\mathcal{T}\{\phi(\zeta)\}=\left[\prod_{n=0}^{\infty}e^{R_{n}[f,\zeta)L_{n}}\phi\right](f(\zeta))
%=\left[e^{R_0L_0}(1+R_1L_1+\ldots)(1+R_2L_2+\ldots)\ldots\phi\right](f(\zeta))
\end{equation}
After expanding the exponentials the infinite product of operators can be rewritten as
\begin{multline}\label{prod}
\prod_{n=0}^{\infty}e^{R_{n}[f,\zeta)L_{n}}\phi\\
=e^{R_0L_0}\left(1+R_1L_1+\frac{1}{2}R_1^2L_1^2+R_2L_2+\ldots\right)\phi
\end{multline}
There are finite terms since any string of generators with a combined descendance
level larger than the descendance level $m$ of $\phi$ (of scaling dimension $\Delta=h+m$) gives zero.
The appearing $R_{n}[f,\zeta)$ coefficients are also known \cite{Gaberdiel}, the first few
being
\begin{align}
&R_0[f,0_{1,2})=\log f'=\log(\pm e^{i\pi d/2}\sin\pi d)\\
&R_1[f,0_{1,2})=f''/2f'=\frac{3+e^{2i\pi d}}{4}\\
&R_2[f,0_{1,2})=Sf/6=\left(\frac{e^{2i\pi d}-1}{4}\right)^2
\end{align}
with the Schwarzian $Sf$. The sign in the first line is crutial and comes form mapping the first and
second planes to the different branches of the square root.
By evaluating (\ref{prod})  it is a matter of algebraic manipulations to find the transformation law for any specific field $\phi$.

\begin{figure}[h]
\begin{centering}
\includegraphics[width=4cm]{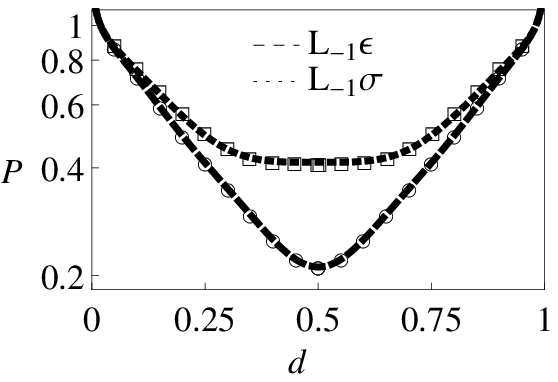}
\includegraphics[width=4cm]{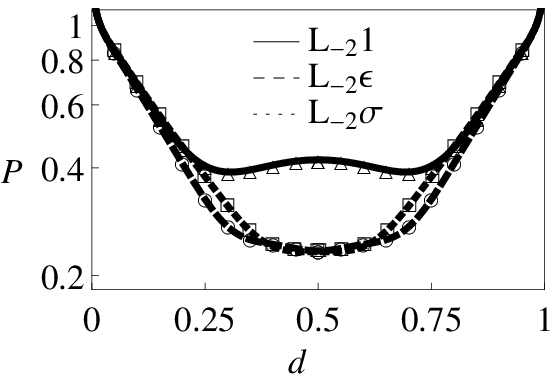}
\par

\includegraphics[width=4cm]{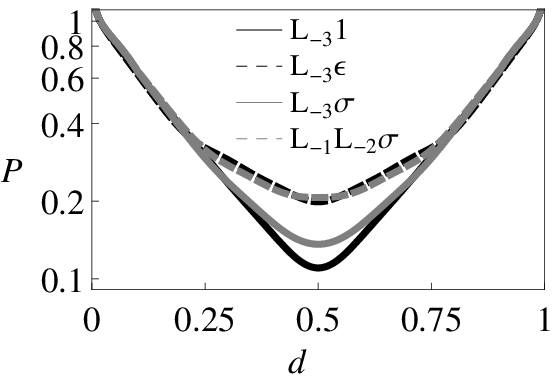}
\includegraphics[width=4cm]{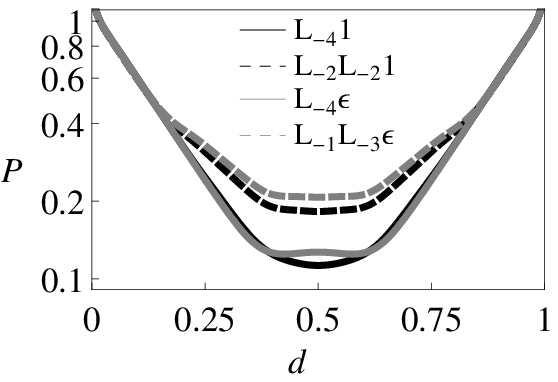}
\par\end{centering}

\caption{(Logarithmic) plots of the subsystem purities of the first few spin-zero
excited states in the Ising universality class ($\Psi=\psi\bar{\psi}$, the legend
shows only the chiral generators $\psi$). We organized the plots
according to the descendance level of the excited states (and the
primaries are not shown). The exponents of the purity in the (first)
exponential domain depend on the excitation energy in a nonlinear
way. The R\'enyi entropy $S_{2}=-\log P$ can also be read off these
plots. For the level one and two states we also show data from the
critical Ising spin chain marked on the plots by points. The agreement
is very convincing and serves as a check of the present framework.
Note, that the scale is arbitrary depending on the regularization
(e.g. system size) and was set so that the exponential decay
would begin around $P=1$.}
\end{figure}

\begin{table}
\scriptsize
\renewcommand{\arraystretch}{1.4}
\begin{center}
    \begin{tabular}{ l  l  l  l llll}
    $\psi$&$1$&$\cos 2\pi d$&$\cos 4\pi d$&$\cos 6\pi d$&$\cos 8\pi d$&$\cos 10\pi d$&$\cos 12\pi d$\\
    \hline
    $1$&1\\
    %$\sigma$&1\\
    $\varepsilon$&$\frac{7}{8}$&$\frac{1}{8}$\\
    $L_{-1}\varepsilon$&$\frac{1558}{2048}$&$\frac{439}{2048}$&$\frac{26}{2048}$&$\frac{25}{2048}$\\
    $L_{-2}1$&$\frac{426347}{524288}$&$\frac{53640}{524288}$&$\frac{38076}{524288}$&$\frac{6200}{524288}$&$\frac{25}{524288}$\\
    $L_{-2}\varepsilon$&$\frac{6085442}{8388608}$&$\frac{1693410}{8388608}$&$\frac{514952}{8388608}$&$\frac{49813}{8388608}$&$\frac{9270}{8388608}$&$\frac{35721}{8388608}$\\
    $L_{-3}1$&$\frac{5569438}{8388608}$&$\frac{2319464}{8388608}$&$\frac{250807}{8388608}$&$\frac{187108}{8388608}$&$\frac{29426}{8388608}$&$\frac{31924}{8388608}$&$\frac{441}{8388608}$\\
    \hline
    \end{tabular}
\end{center}
\caption{Square roots of the scaling functions $F_\Psi$ for some spin zero states
in the Ising CFT given as sums of cosines,
$\sqrt{F_\Psi(d)}=\sum_{n=0}^{n_\text{max}} c_n \cos(2\pi n d)$.}
\end{table}

After the transformation one is left with a sum of
four-point functions of descendant fields in the form
\begin{equation}
F_\Psi=\sum_{abcd}c_{abcd}\left\langle \phi_{a}(\zeta_{1})\phi_{b}(\zeta_{2})\phi_{c}(\zeta_{3})\phi_{d}(\zeta_{4})\right\rangle _{\mathbb{C}}
\end{equation}
The exact evaluation of such n-point functions is in principle
known and it can be obtained from the four point function of the associated
primary fields by acting on it with an appropriate differential operator.
The simplest example is the four-point function of three primaries
and one descendant, e.g. 
\begin{multline}
\langle L_{-n}\Phi_{1}(x_{1})\Phi_{2}(x_{2})\Phi_{3}(x_{3})\Phi_{4}(x_{4})\rangle\\
=\sum_{i=2}^{4}\left\{ \frac{(n-1)h_{i}}{(x_{i}-x_{1})^{n}}-\frac{1}{(x_{i}-x_{1})^{n-1}}\frac{\partial}{\partial x_{i}}\right\} \\
\langle\Phi_{1}(x_{1})\Phi_{2}(x_{2})\Phi_{3}(x_{3})\Phi_{4}(x_{4})\rangle%\\
%\equiv \mathcal{D}^{L_{-n},1,1,1}(x_1,x_2,x_3,x_4)\langle\Phi_{1}(x_{1})\Phi_{2}(x_{2})\Phi_{3}(x_{3})\Phi_{4}(x_{4})\rangle
\end{multline}
where the particular differential operator is easily inferred from
the conformal Ward identities. To find the differential operator 
%$\mathcal{D}^{S_1,S_2,S_3,S_4}(x_1,x_2,x_3,x_4)$ 
in the general case, i.e. for arbitrary strings of Virasoro generators,
we rewrote the generators as contour integrals
\begin{equation}
L_{n}\phi(x)=\frac{1}{2\pi i}\oint_{x}d\zeta(\zeta-x)^{n+1}T(\zeta)\phi(x)
\end{equation}
and we deformed the contours successively from one operator insertion
point to the others, back and forth. When doing the deformations the
generators $\{L_{n}\}_{n\geq-1}$ are generated which can be seen by
expanding $(\zeta_1-x)^{n+1}$ in powers of $(\zeta_2-x)$. Using this technique it is
possible to reduce any $n$-point function into a sum of ones that
involve only the generator $L_{-1}$, which is equivalent to partial differentiation. 
The residual four-point functions of primaries can be obtained exactly for instance by means
of the Coulomb gas construction \cite{Difrancesco} as a sum of chiral$\times$antichiral
products of hypergeometric functions (conformal blocks) \cite{foot1}.
All these computations become extremely cumbersome, producing a large number
of terms for already the simplest descendant states. Therefore we algorithmized
the transformation and the computation of the differential operator and did them
still analytically assisted by computer. Further details of this approach will be discussed in a later
publication \cite{later}, where we shall present a different application
of excited state R\'enyi entopies.

\begin{figure}[b!]
\begin{centering}
\includegraphics[width=4cm]{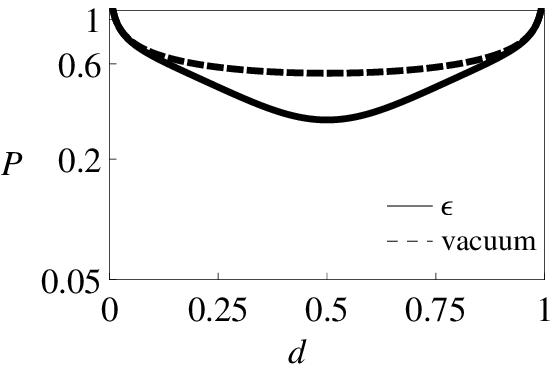}
\includegraphics[width=4cm]{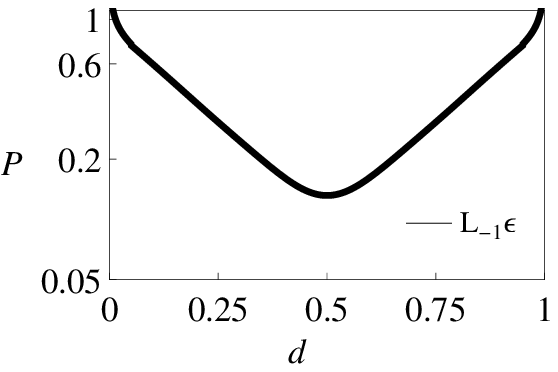}
\par
\includegraphics[width=4cm]{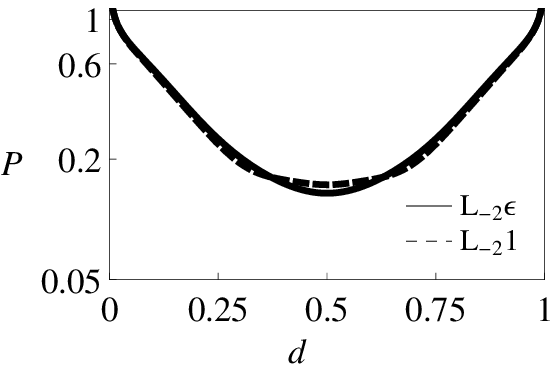}
\includegraphics[width=4cm]{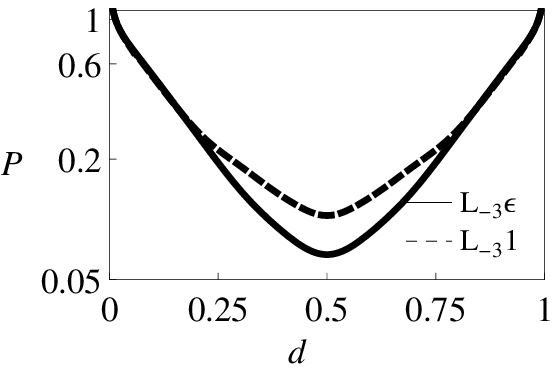}
\end{centering}

\caption{(Logarithmic) plots of the subsystem purities of the first few spin-zero
excited states from the identity and energy towers in the Potts universality
class.%up to descendence level 3.
}
\end{figure}

\paragraph*{Results and discussion}

In Figs. 1 and 2 we show results for the subsystem purity for excited
states in the Ising ($c=1/2$) and the three state Potts ($c=4/5$) models.
For the former case we also checked the results for energetically non-degenerate states in the zero momentum sector (where the identification with the corresponding excited states in the CFT is straightforward by comparing excitation energies) by performing calculations on the critical Ising spin chain (of 200 spins) and found perfect agreement (see Fig. 1).\cite{foot2} 
In Tab. 1 we show some of the exact $F_\Psi$ functions in the Ising model.

In addition to the general characteristics already discussed we see
that for certain states one can
identify multiple decay exponents important in different domains of the
subpurity. While the first exponent depends only
on the excitation energy (\ref{eq:Sexc}) the further exponents are
different for degenerate states. Based on this observation we can
also define a purity spectrum $\{p_{1},p_{2},\ldots,p_{n}\}$ consisting
of all the exponents relative to the specific state. Indeed, it would
be very interesting to better understand this form of the purity through
an explicit calculation of the exponents in terms of the conformal
data and to understand their physical meaning. In fact, a similar
structure was found in \cite{excspinchain} for the massive XY spin chain
and it was suggested that (at least some) excited states can be reinterpreted
as ground states of systems of coupled spin chains where applying
the area law leads to an extensive behavior
for the entanglement entropy with slope-changes depending on the specific
state.

\paragraph*{Acknowledgments.}

I am grateful to G. Tak\'acs for help and encouragement at various stages
of this project. I also thank G. Mussardo, I. D. Rodriguez, F. Franchini
and G. Sierra for valuable discussions.

%\newpage


\begin{thebibliography}{10}
\bibitem{exceptions} Excited states where Eqs.  (\ref{Ssmalld}-\ref{eq:Sexc}) are expected to break down include ground states of different Hamiltonians but apparently also those states that are true ground states only in the thermodynamic limit and are equivalent to the finite volume ground state under some global symmetry. Examples include the lowest-lying excitation in the Ising model and states generated by vertex operators in the massless free boson theory \cite{Sierra} (corresponding to $Z_2$ and $U(1)\times U(1)$ symmetries, respectively). While intuitively we understand that such states should be equivalent also with respect to entanglement, at least in the thermodynamic limit, a complete explanation is missing.
\bibitem{foot1}An alternative, numerical way of calculating four-point
functions of descendents is obtained by inserting a resolution of
identity and summing up three-point functions. However, in the course
of our calculations we found that we need the exact four-point functions
to get reliable results for the subpurity.
\bibitem{foot2}The subpurity of states in the Ising spin chain,
with the Hamiltonian $H=-\frac{1}{2}\sum_{j=1}^{N}\sigma_{j}^{x}\sigma_{j+1}^{x}+J\sigma_{j}^{z}$
(being critical at $J=1$), can be straightforwardly calculated in
terms of Majorana fermions by the technique presented in \cite{Latorre}
and the Appendix of \cite{SierraJStat}.
\end{thebibliography}
\end{document}